\documentclass[twocolumn,preprintnumbers,amsmath,amssymb,superscriptaddress]{revtex4}
\usepackage{graphicx}% Include figure files
\usepackage{dcolumn}% Align table columns on decimal point
\usepackage{bm}% bold math

\begin{document}
\title{The two-step photoexcitation mechanism in a-Se}
\author{J. Berashevich, A. Mishchenko, and A. Reznik}
\affiliation{Thunder Bay Regional Research Institute, 290 Munro St., Thunder Bay, ON, P7A 7T1, Canada}
\affiliation{Department of Physics, Lakehead University, 955 Oliver Road, Thunder Bay, ON, P7B 5E1}

\begin{abstract}
The first-principal simulations are applied to study a photo-induced metastability in
amorphous selenium (a-Se) and the contribution of the valence-alteration pair (VAP) defects in this process.
The VAP defect is confirmed to be the equilibrium defect; it minimizes the destabilizing
interaction between adjacent Se chains induced by dis-orientation of the lone-pair (LP) electrons,
and thus relieves a tension in a system. The excitation of LP electrons is proposed to be
described by two coexisting processes, namely, single and double electron excitations.
Both processes have been found to form defect states in the band gap and to cause the experimentally
observed photo-darkening and photo-volume expansion, however, only double electron excitation is
capable to trigger bond rearrangement and structural transformation. Lattice relaxation,
which follows bond rearrangement occurs with characteristic energy of -0.9$\pm$0.3 eV and
promotes formation of energetically favorable VAP defects or crystalline inclusions thus
ultimately stimulating the photo-induced crystallization. In addition, photo-induced
crystallization has been directly simulated in a system with an increased crystalline order.
\end{abstract}

\maketitle
\section{Introduction}
The progressively expanding area of applications of chalcogenide glasses (from medical imaging detectors
to phase-change memories) supports an irrefutable interest in their properties, one of which is
photoinduced transformation - a very unique phenomenon that is a special feature of chalcogenide glasses including amorphous Se.
The transformation is known to be triggered by optical or x-ray irradiation and is associated
with a creation of defects \cite{1,2,3}. 
The most studied defect in amorphous Se is a valence-alteration pair (VAP) \cite{1,2,3,4,5}
which belongs to "equilibrium" type of defects \cite{4} meaning that VAP exists in well relaxed
structure prior to photoexcitation. However, photoexcitation 
is assumed to enhance a concentration of the VAP defects by initiating the bond breaking reaction:
2C$_2^0\leftrightarrow$C$_3^+$+C$_1^-$, where C$_2^0$
is a selenium atom in its normal two-fold configuration, 
C$_3^+$ is a positively charged three-fold configuration, and C$_1^-$
is a negatively charged one-fold coordinated Se atom \cite{1}. 
The direct evidence of the valence alteration in a-Se is experimentally
detected unpaired electrons which appear under light illumination \cite{2}. 
It is believed that the VAP defects are
responsible for many of the photoinduced effects including photodarkening (PD) \cite{6,7,8}, 
photoinduced crystallization \cite{7} and volume expansion \cite{13,14,15} as they induce
the sub-band gap defect states \cite{9,10,11,19} .
Although a link between these effects is not confirmed with direct experiments, 
they all have similar kinetics  and are shown to be thermally activated
processes with similar activation energies around $\sim$0.9 eV \cite{6,7} - 
the value which was suggested for the VAP defect relaxation \cite{12}.

In order to get a better insight into the nature of the photoinduced effects
and contribution of the VAP defects in it, we model the entire process of
photoexcitation with help of the first-principles methods.
First, we have classified all possible VAP defects in a-Se network,
their energetics and positions in the band gap.
Then, we have simulated a photoexcitation from an "ideal" system (consisting 
of strictly two-fold coordinated Se atoms) and a system containing the VAP defect.
We show that in both systems the excitation of two electrons from the
lone pair (LP) state may trigger a formation of a
dynamic bond with the characteristics energy -0.9$\pm$0.3 eV through
the photo-induced relaxation of the immediate neighborhood. 
In post-excitation regime, this dynamic bond is broken which then
induces the bond rearrangement followed by a creation either of the
VAP defect or the crystalline inclusion.
In contrast, removal of just one electron from the LP state does not
generate any significant lattice relaxation and, therefore, 
does not lead to a permanent structural transformation.
The simulated photo-induced transformations show good correlation with 
the PD kinetics measured under the red light illumination. 
The comparison with the experimental results suggests that 
the slow component in the kinetics of the photodarkening relaxation can be linked to the process which
involves the bond rearrangements (i.e. when two electrons are photoexcited from the LP state)
with the experimentally found activation energy E$_B$=0.8$\pm$0.1 eV.
In contrast, the fast component in the photodarkening relaxation corresponds to a process 
which does not involve any significant lattice relaxation that is 
observed in simulations of a single electron excitation.

\section{Theoretical methods}
For our study, we created the model compounds of amorphous
Se consisting of 25 atoms and 50 atoms. For the atomistic simulations, the WIEN2k package 
\cite{16} with implementation of the Perdew-Burke-Ernzerhof parametrization of the
generalized gradient approximation \cite{17} has been used.
The optimization procedure was carried based on minimization of forces
for which the product of atomic sphere radius (1.8 Bohr centered at the nucleus of the individual atoms)
and plane-wave cutoff in $k$-space was set to 7, 
while the Brillouin zone of a supercell was covered by the 4$\times$4$\times$4 Monkhorst-Pack mesh.
The energy separating core and valence electrons was set to -6.0 Ryd. The force tolerance 0.5
mRyd/Bohr in combination with tight convergence limits (energy convergence 0.0001 Ryd,
force convergence 0.1 mRyd/Bohr and charge convergence 0.001 e) have been applied. 

To study the fundamental properties of a-Se including the formation of the VAP defects
and photo-induced structural transformations, the well amorphized a-Se placed in cubic supercell
($\alpha$=90$^{\circ}$, $\beta$=90$^{\circ}$, $\gamma$=90$^{\circ}$) with periodic boundary conditions was used 
(the size of the cubic supercell consisting of 25 atoms was 8.7\AA$\times$8.7\AA$\times$9.9\AA).
In order to insure that the inter-cell defect interaction does not effect the energetics of the VAP defects, 
the results have been verified with a bigger supercell consisting of 50 atoms.
The size of the Monkhorst-Pack mesh was adjusted 
to the supercell size, i.e. the mesh was appropriately reduced 
as the supercell was enlarged.
In the created model compounds, the bond length between two-fold sites was found
to vary in a range of 2.30-2.47 \AA\ that agrees well with an average bond length in a-Se of 2.37 \AA\ \cite{12,13}.
The bond angles along the chains in the well amorphized system have shown to vary from 98$^{\circ}$ to 107$^{\circ}$
against 102$^{\circ}$ in trigonal Se (t-Se). For the simulations of the photo-induced crystallization,
the system of the increased crystalline order has been created.
To achieve crystallization, the model system has been assigned to 
recognize the symmetry of trigonal Se through application of the hexagonal
lattice parameters to supercell ($\alpha$=90$^{\circ}$, $\beta$=90$^{\circ}$, $\gamma$=120$^{\circ}$).

\section{The VAP defects and their classification}
We start our analysis from the "ideal" amorphous system containing only two-fold
coordinated Se atoms, each having two electrons on the $p$-bonding orbital,
two empty quantum states on the $p$-anti-bonding orbital and two electrons on the nonbonding LP orbital.
The properties of such "ideal" network are governed by the lone pairs: by pushing
the shared orbitals apart, they define the bond angle and torsion angle.
The crystalline systems, i.e. trigonal and $\alpha$-monoclinic Se, are built
in a way to minimize interatomic interactions of the LP electrons. 
In amorphous network, LPs are disoriented that causes
tension between chains and as a result, the total energy 
is larger by $\Delta$=0.08 eV per atom than that
in t-Se due to the destabilizing interactions
between chains. The charge exchange between Se atoms in amorphous network
(up to $\pm$0.5 $\bar{e}$ per atom) is an attempt to reduce the chain tension.
Naturally, the crystalline-like inclusions suppress the tension. We estimated that formation of a Se$_7$ ring
results in energy lowering by $\Delta$=-0.02 eV per atom. 

With respect to the electronic properties, disorientation of LPs
causes formation of the band tails. The edge of the valence band (E$_V$)
generated by the lone pairs is found to be smeared by the broad tails as defined by
$\Delta$E$_V$=E$_V$+0.45 eV.
Because bottom of the conduction band (E$_C$) is formed by $p$-antibonding orbitals, 
the effect of LP disorientation is less pronounced there \cite{18}
giving a rise to the band tails of a size of $\Delta$E$_C$=E$_C$-0.16 eV. 
This simulated asymmetry in the band tails has also been observed experimentally \cite{9,10}.
It has to be mentioned that an inclusion of a Se$_7$ ring locally improves the 
LPs orientation resulting in total band tails shrinkage
by $\sim$0.3 eV with most pronounced effect near the valence band.

\begin{figure*}
\includegraphics[scale=0.35]{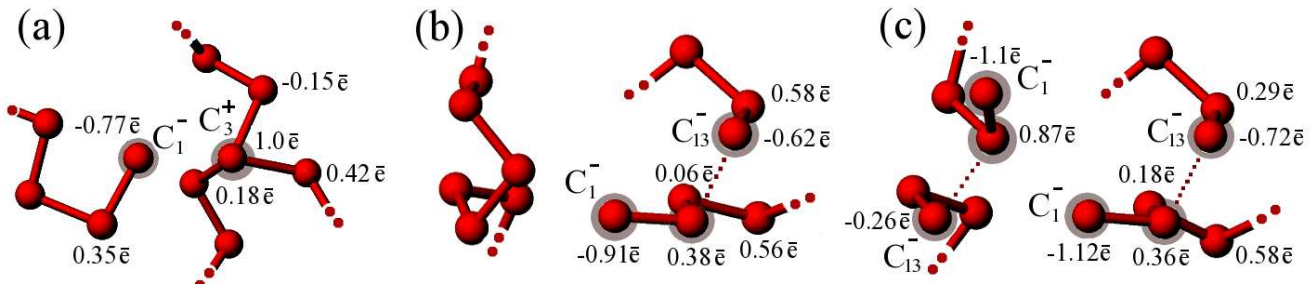}
\caption{\label{fig:fig1} The VAP defects with charge distribution
(a) two chains VAP defect; (b) single IVAP defect; (c) pair of IVAP defects.}
\end{figure*}

In addition to crystalline-like inclusions, another way to resolve the destabilizing
interactions between two chains is chain discontinuation C$_2^0\rightarrow$C$_1^0$+C$_1^0$. 
When chain breaks, one C$_1^0$ end remains as the separated site, 
while another connects to two-fold C$_2^0$ site of another chain thus becoming three-fold C$_3^0$ 
(see defect structure in Fig. 1 (a)). This process is followed by the charge exchange reaction 
C$_3^0$+C$_1^0\rightarrow$C$_3^+$+C$_1^-$ (see charge distribution in Fig.~\ref{fig:fig1} a)
and formation of two chains VAP defect \cite{2,3}. 
We have found that the total energy lowering with this defect formation is $\Delta$=-0.05 eV per atom 
(since this value is calculated per all atoms in a supercell, it is underestimated for atoms involved in the VAP formation). 
Therefore, we numerically confirmed that the two chains VAP defect is very efficient in relieving tension between chains despite
a strong dipole formed between the C$_3^+$ and C$_1^-$ sites.
With respect to the defect structure, the C$_3^+$ and C$_1^-$ sites are separated
by a distance 3.45 \AA, while the bond length around C$_3^+$ is slightly
elongated to 2.4-2.6 \AA\ due to the bond iconicity. 

When broken C$_1^0$ end connects to the same chain, another
type of the VAP defect emerges (see Fig.~\ref{fig:fig1} b) \cite{5}
which is named in literature as "intimate" VAP defect (IVAP) \cite{5}.
We have found that the IVAP defect is stable only in the intimate configuration. 
Mechanism of formation of the IVAP defect is the same as for VAP (Fig.~\ref{fig:fig1} a), 
one broken end remains detached while second attempts
to convert into three-fold site denoted in Fig.~\ref{fig:fig1} b as C$_{13}^-$.
Because the induced charge redistribution is restricted to a single chain,
the originated strong dipole keeps apart the
C$_{13}^-$ and C$_{2}^+$ sites at distance $d_{\operatorname{C}_{13}^- -\operatorname{C}_{2}^+}$=2.82 \AA\ 
(see a dashed line in Fig.~\ref{fig:fig1} b).
Moreover, a strong dipole moment masks the stress relieving effect 
and as a result, no decrease in energy is detected with this defect formation.
However, when two IVAP defects appear in a pair (see Fig.~\ref{fig:fig1} c showing two IVAP defects separated by a distance $\sim$3.8 \AA),
the interaction between defects induces the partial compensation of their dipoles. It allows the bond shortening between
C$_{13}^-$ and C$_{2}^+$ sites to $d_{\operatorname{C}_{13}^- -\operatorname{C}_{2}^+}$=2.78 \AA\ and
uncovers the stress relieving effect resulting in the energy lowering by $\Delta$=-0.04 eV per atom. 
The IVAP pair is found here for the first time.
Since the total energy decreases with pair formation, this
pair configuration should dominate over single IVAP.

Overall, the "ideal" system has found to be the highest in energy.
Therefore, the VAP defects or crystalline inclusions such as the Se$_8$, Se$_7$ and Se$_6$ rings \cite{20}
are thermodynamic defects \cite{4} which formation reduce the stress originated from LPs dis-orientation.
As a result, in a-Se grown samples,
there will always be interplay between concentration of the rings and the VAP defects.
The preference would be given to formation of the VAP defects (Fig.~\ref{fig:fig1} a)
as they are most efficient in relieving tension between chains
reflected by the greatest energy reduction $\Delta$=-0.05 eV per atom.
This explains the low concentration of the rings observed experimentally 
(fraction of atomic rings is 5$\%$-10$\%$ \cite{21,22}).
In regions under enhanced stress such as interfaces,
the higher concentration of the tension relieving centers is expected.

The VAP defects have direct impact on the electronic properties of a-Se.
According to our calculations, the defect states generated by C$_{3}^+$ 
appear below the conduction band at E$_C$-0.33 eV for the two chains
VAP; at E$_C$-0.43 eV for the single IVAP; and at E$_C$-0.61 eV for the IVAP pair. 
The energetic positions of the defect states induced by C$_{3}^+$ agrees
well with the experimental observation showing two peaks in
the density of states at ~0.30 eV and 0.45$\div$0.50 eV below E$_C$ \cite{10}
that is also consistent with the positions of the charged centers in Ref.\cite{11}.
The C$_{1}^-$ states are located above the valence band at
E$_V$+0.34 eV for the VAP and the single IVAP defects, and at E$_V$+0.23 eV for the IVAP pair. 
Therefore, the C$_{1}^-$ states appear inside the valence band tails defined by $\Delta$E$_V$=$E_V$+0.45 eV. 
Depending on the concentration of the VAP, single IVAP and IVAP pairs, they 
may influence the density of state distribution causing characteristic features 
at about 0.23 eV and 0.34 eV above the valence band as observed in some experiments \cite{9}. 

Therefore, our finding confirms that the VAP defects are the thermodynamic or the "equilibrium" type 
as suggested in literature \cite{4}, i. e. they should be present at high concentration
in the well-relaxed structures even prior to light excitation. 
Since the formation of the VAP defects and the crystalline
inclusions in a form of the Se$_7$ ring are energetically favorable, 
their concentration should grow under light illumination.
Hence, our next step is a simulation of the photoinduced changes in amorphous Se network.

\section{Photo-induced lattice relaxation}
\begin{figure}
\includegraphics[scale=0.80]{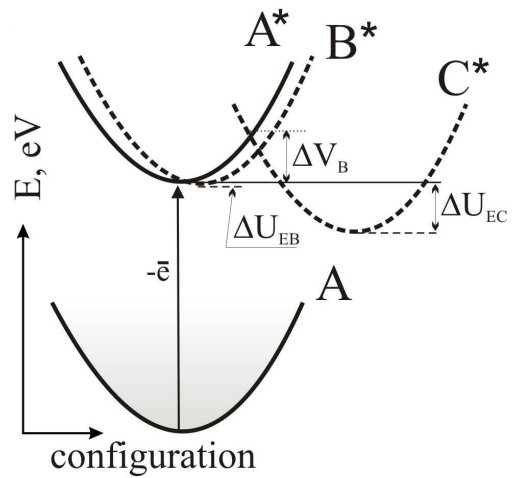}
\caption{\label{fig:fig2}Potential diagram describing an excitation of the A state and 
contribution of the photo-induced lattice relaxation into stabilization of the excited state A$^*$.}
\end{figure}

The photoexcitation process can be described with potential diagram presented in Fig.~\ref{fig:fig2}. 
The excitation of electrons brings the system from a ground state A to an excited state A$^*$.
We assume that an electron excited into the conduction band leaves behind the hole in the valence band tails.
The state in the valence band tails missing one or two electrons induces the lattice
relaxation of the immediate neighborhood needed for its stabilization.
As a result, the excited A$^*$ state converts either into B$^*$ or C$^*$ state. 
The B$^*$ state accounts for the insignificant lattice relaxation
involving no bond rearrangements as denoted by small energy lowering $\Delta$U$_{EB}$. 
In contrast, the C$^*$ state occurs with bond rearrangements reflected in large $\Delta$U$_{EC}$.
The A$^*$$\rightarrow$C$^*$ transfer requires overcoming the potential barrier $\Delta$V$_B$. 
When missing electrons return, only the B$^*$ state would be able to relax into the original ground state A. 

The sub-bandgap excitation occurs primarily from the top of the valence band
formed by the LP electrons: for the "ideal" system an excitation involves the partially charged 
C$_{2}^-$  sites, while for the VAP defects the C$_{1}^-$ sites are those who participate. 
Upon excitation, the LP occupancy is reduced by one electron (LP$^2-\bar{e}\rightarrow$LP$^1$) 
and the LP$^1$ state with unpaired electron is shifted deeper into the bandgap by $\sim$0.2 eV
due to an alteration in the exchange interaction upon removal of an electron. 
The LP$^1$ states induce the PD effect and are detected experimentally due to a presence of the unpaired electrons \cite{2}. 
The electron density of the LP$^1$ state is delocalized over several sites:
our calculations show that it involves up to 5 sites for the two chains VAP defect. 
The relaxation of the immediate neighborhood around the LP$^1$ state required
for its stabilization is found to be accompanied by the bond shortening/elongation
occurring with characteristic energy $\Delta$U$_{EB}<$-0.1 eV that corresponds to the B$^*$ state in Fig.~\ref{fig:fig2}. 
The bond shortening is observed for the IVAP defects (see Fig.~\ref{fig:fig1} b,c). 
The $d_{\operatorname{C}_{13}^- -\operatorname{C}_{2}^+}$ bond shortens by 0.1 \AA\ that agrees with previous findings \cite{13}. 
In contrast, the double chain VAP defect shows a significant increase
in a distance between the C$_{3}^+$ and C$_{1}^-$ sites by 0.25 \AA. 
Because the concentration of the double chain VAP defects should dominate over other defects, 
an increase in a separation between C$_{3}^+$ and C$_{1}^-$ sites would
govern volume expansion observed experimentally \cite{14}. 
Since the bond elongation occurs with the characteristic energy $\Delta$U$_{EB}$, it is not an immediate process. 
Therefore, the volume expansion caused by LP$^1$ states at the double chain VAP defect
is expected to be delayed from the PD onset that agrees with the experimental observations \cite{14}. 
When the missing electron returns (LP$^1+\bar{e}\rightarrow$LP$^2$), 
all systems have been found to come back to the original state A. 

It should be noted that for the formation of the photo-induced robust dynamic bond 
known as the C$_3$-C$_3$ defect \cite{2,3}, 
an excitation of two electrons in a vicinity of a single site is required that is 
also simulated here as LP$^2-2\bar{e}\rightarrow$LP$^0$ process 
(LP$^0$ appears above the LP$^1$ position in the band gap).
Because excitation of two electrons is unlikely to occur simultaneously,
two possible scenarios are considered. (i) LP$^1-\bar{e}\rightarrow$LP$^0$: 
inequality in the charge distribution generated by LP$^1$ promotes
excitation of the second electron from the vicinity of the same site. 
(ii) LP$^1$+LP$^1$$\rightarrow$LP$^0$+LP$^2$: two interacting LP$^1$ states undergo charge exchange.
The latter process becomes feasible when the energy lowering
due to the lattice relaxation needed for stabilization of the LP$^0$ state
is greater than 2$\Delta$U$_{EB}$ accounted for two interacting LP$^1$ states. 

The double electron excitation has been simulated by successive removal of two electrons from the system.
For the "ideal" two-fold coordinated system and the IVAP pair defect (Fig.~\ref{fig:fig1} c), 
the significant lattice relaxation with the characteristics energy
$\Delta$U$_{EC}$=-0.9$\pm$0.3 eV is followed by the bond rearrangements (the C$^*$ state in Fig. 2).
In the "ideal" two-fold coordinated system, the C$_3$-C$_3$ dynamic bond occurs
through crosslinking between two chains \cite{2,3}. 
The defect state induced by the C$_3$-C$_3$ dynamic bond appears close to the midgap.
Following lattice relaxation, two unequally populated
LP$^1$ sites formed upon excitation are converted into the LP$^0$ and LP$^2$ states,
i.e. the charge transfer reaction LP$^1$ + LP$^1$$\rightarrow$LP$^0$ + LP$^2$
is initiated since system meets the requirement $|\Delta$U$_{EC}|>2|\Delta$U$_{EB}|$. 
For the IVAP pair defect, the double excitation has been found to induce
bond rearrangements followed by formation of the single C$_3$ defect with three equivalent bonds
2.4-2.5 \AA\ (the defect state appears closer to the conduction band).
To reach the final configuration,
the lattice relaxation requires overcoming the potential barrier $\Delta$V$_{B}\sim$0.8 eV.
It correlates well with the potential diagram presented in Fig.~\ref{fig:fig2} for A$^*$$\rightarrow$C$^*$ transition.
However, some configurations have shown to be resistant to the
double electron excitation because destabilization 
induced by the hole localized in the valence band tails is not enough to overcome the potential barrier $\Delta$V$_{B}$.
Thus, for the two chains VAP and the single IVAP defects, the B$^*$ 
state with characteristic relaxation energy $\Delta$U$_{EB}<$ -0.1 eV occurs predominantly.

When two missing electrons are returned into the system, the C$^*$ state
becomes unstable and its stabilization requires breaking the dynamic
or other bonds that depends on whether the dynamic bond is strongly stabilized
at the excited state or not. The process of bond breaking and switching in
the post-excitation regime is found to occur with the characteristic energy 
-0.9$\pm$0.3 eV which is similar to $\Delta$U$_{EC}$. In this case, the final configuration
is different from the original ground state A (Fig.~\ref{fig:fig2}).
The "ideal" system has been converted into one containing the S$_7$ ring; decrease
in the total energy as compared with the initial configuration is found to be 0.49 eV. 
The IVAP pair (Fig.~\ref{fig:fig1} c) is converted into the double chain VAP defect (Fig.~\ref{fig:fig1} a)
accompanied by the lowering in energy by 0.28 eV.
Therefore, we conclude that following the sub-bandgap 
illumination which involves excitation of the LP electrons, the system goes
into the most stable state characterized by the increased concentration of the S$_7$ ring
and/or the VAP defects. 

\begin{figure}
\includegraphics[scale=0.19]{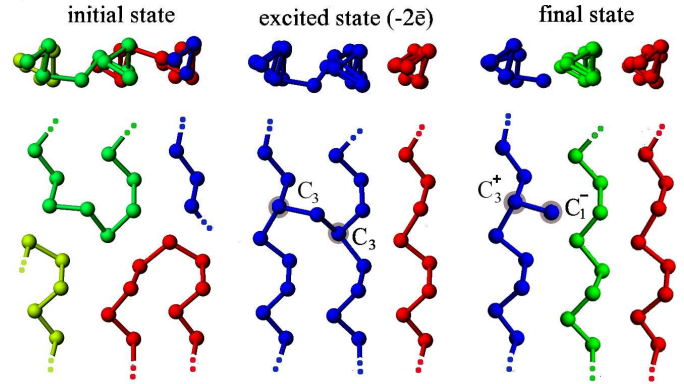}
\caption{\label{fig:fig3} Colour on-line. The photo-induced crystallization of the increased
crystalline order system. The final state is t-Se containing the IVAP defect.}
\end{figure}

Since the double electron excitation is sufficient to induce the bond rearrangements,
we attempt to generate the photocrystallization in the increased crystalline
order system as shown in Fig.~\ref{fig:fig3}.
The electron excitation is found to induce the formation of the standard C$_3$-C$_3$ defect 
followed by chain straightening (see the excited state in Fig.~\ref{fig:fig3}).
In post-excitation regime, the complete conversion to t-Se is achieved.
Because of the increased crystalline order of the initial system,
the characteristics energies describing the lattice relaxation is found to be
$\Delta$U$_{EC}=$-0.5$\pm$0.3 eV that is lower than that for the well amorphized systems.
Therefore, we suggest that experimentally observed crystallization of the a-Se samples 
under sub-bandgap illumination \cite{7,23} should be linked to 
the double electron excitation process. 

\section{The structural transformations in PD kinetics}
In order to verify that both single and double electron excitations
are plausible processes involved in photoexcitation, the kinetics of PD
in a temperature range between room temperature and 40 $^{\circ}$C have been studied experimentally
(the upper temperature limit is dictated by a-Se glass transition temperature).
The PD experiments were carried out on 15$\mu$m thick stabilized a-Se layer (with 0.5$\%$ of As)
using standard two red laser beams setup (655 nm), where a powerful pump beam (150 mW/cm$^2$)
was used to produce photodarkening while less powerful probing beam (0.29 mW/cm$^2$)
monitored changes in transmission of light T.
The detailed descriptions of the experimental apparatus can be found in the previous work \cite{6}. 

\begin{figure}
\includegraphics[scale=0.35]{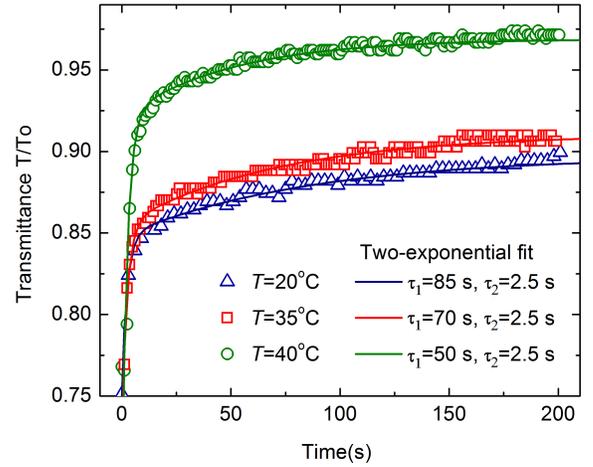}
\caption{\label{fig:fig4} Colour on-line. Symbols: the experimental results 
on the PD relaxation in a-Se for three selected temperatures namely,  
restoration of the probing beam transmission after switching off illumination
(rest period). Solid lines: the fitted theoretical functions.}
\end{figure}

The kinetics of PD were studied by periodically exposing the a-Se sample
to the pumping beam for 200 s separated by 200 s of rest.
During those cycles, the probing beam transmission, T, is continuously
monitored and the relative changes compared to the original transmission of light,
T/T$_0$, were then calculated. The restoration of the transmission during
the rest period at room temperature and selected elevated temperatures is shown in Fig.~\ref{fig:fig4}
and can be modeled by the double-exponential decay yielding two
characteristic time constants: $\tau_1$ and $\tau_2$. At room temperature,
the characteristic relaxation times $\tau_1$ and $\tau_2$ derived from
fitting of the resting cycles are 85 s and 2.5 s, respectively, after averaging over several cycles.
Remarkably, $\tau_1$ decreases with increase in temperature showing the Arrhenius dependence:
\begin{equation}
\tau_1=\nu_0^{-1}exp(\frac{E_B}{k_B\Theta})
\label{eq:rate1}
\end{equation}
where $k_B$ is the Boltzmann constant, $\Theta$ is the temperature and $\nu_0$ is the attempt-to-escape frequency.
It yields the activation energy E$_B$=0.8$\pm$0.1 eV ($\nu_0$=$2\cdot$10$^{11}$s$^{-1}$).
In contrast, $\tau_2$ remains temperature independent for all applied temperatures. 

This clearly demonstrates that relaxation of PD involves two distinct processes:
one that has the lattice relaxation in its origin and requires activation over a
potential barrier with characteristic time $\tau_1$, and another process which does not involve
the lattice relaxation and occurs much faster with $\tau_2$. 
The remarkable similarities between the magnitude of the activation energy E$_B$=0.8$\pm$0.1 eV
found in PD experiment and $\Delta$U$_{EC}$=-0.9$\pm$0.3 eV required for the bond rearrangements, 
makes us to believe that this slow component in PD describes the
process of structural relaxation with the bond rearrangements. Therefore, a feasibility of the process of
double electron excitation accompanied by formation of the LP$^0$ states is confirmed by slow component of the PD relaxation. 
Taking into the account two plausible scenarios for the LP$^0$ sites formation
(LP$^1-\bar{e}$$\rightarrow$LP$^0$ or LP$^1$+LP$^1$$\rightarrow$LP$^0$ + LP$^2$),
we suggest that concentration of the 
LP$^0$ states should depend non-linearly on the rate of excitation. 

In contrast, the temperature independence of the fast component 
($\tau_2$=2.5 s) is the direct evidence of the relaxation process that does not involve bond rearrangements.
It is consistent with very weak lattice relaxation $\Delta$U$_{EB}<$-0.1 eV
involved in stabilization of the LP$^1$ states generated by the single
electron excitation. The experimental detection of the unpaired electrons \cite{2} which can be associated 
only with the LP$^1$ states is further 
evidence of the single-electron excitation. 

\section{Conclusions}
The first-principal methods are applied to simulate the photoinduced
structural transformations in a-Se network that has allowed for the compilation of
the complete picture of the photo-induced changes.
Our calculations suggest two distinct mechanisms of the photo-excitation, i.e. the single electron and double 
electron excitation, which are followed by different lattice relaxation processes. 
The excitation of a single electron from LP$^2$ leaves behind an unpaired electron and shifts
the non-bonding LP$^1$ state towards the midgap thus inducing the photodarkening effect.
This process manifests itself in the fast component of the relaxation of the photodarkening 
which is characterized by short and temperature independent characteristic time constant $\tau_2$=2.5 s. 
The fact that $\tau_2$ is temperature independent suggests that no bond rearrangement is involved in 
the lattice relaxation. 
Since the hole localized at the LP$^1$ site does not trigger the bond rearrangements,
the single electron excitation can not be accounted for the photo-induced crystallization. 
However, it is responsible for the photo-induced volume expansion \cite{14} 
as the LP$^1$ states appearing at the double 
chain VAP defect
increase a separation between the C$_{3}^+$ and C$_{1}^-$ sites. 

In order to trigger the structural transformation in the a-Se network,
excitation of two electrons from the vicinity of the same LP is required.
The characteristics energy of the lattice relaxation following the
formation of the C$_3$-C$_3$ defect is found to be $\Delta$U$_{EC}$=-0.9$\pm$0.3 eV.
This process is reflected in the slower component of the PD relaxation described
by the temperature dependent time constant $\tau_1$=85 s yielding the activation energy of
E$_B$=0.8$\pm$0.1 eV. The similarity between the characteristics energy describing
the lattice transformation $\Delta$U$_{EC}$ and the activation energy E$_B$ in the PD kinetics
suggests their same origin.
In addition, the application of the double electron excitation to the system of
the increased crystalline order has found to induce photocrystallization. \\

\section{Acknowledgement}
The computational facilities have been acquired through membership in
Shared Hierarchical Academic Research Computing Network (SHARCNET:www.sharcnet.ca)
and Compute/Calcul Canada. Authors are also thankful to Dr. Safa Kasap for stimulating discussions and 
to Dr. O. Rubel for sharing his computational clusters.
Financial support of Ontario Ministry of Research and Innovation through a
Research Excellence Program Ontario is highly acknowledged.

\end{document}